# Mixed reality hologram slicer (mxdR-HS): a marker-less tangible user interface for interactive holographic volume visualization

Hoijoon Jung[1], Younhyun Jung[1,2*], Michael Fulham[3,1] and Jinman Kim[1]

[1*] School of Computer Science, The University of Sydney, Sydney, 2006, NSW, Australia.

[2] School of Computing, Gachon University, Seongnam-si, 13120, Gyeonggi-do, South Korea.

[3] Department of Molecular Imaging, Royal Prince Alfred Hospital, Sydney, 2050, NSW, Australia.

*Corresponding author. E-mail: yjun6175@uni.sydney.edu.au;

Contributing authors: hoijoon.jung@sydney.edu.au; michael.fulham@sydney.edu.au; jinman.kim@sydney.edu.au;

**Abstract**

Mixed reality head-mounted displays (mxdR-HMD) have the potential to visualize volumetric medical imaging data in holograms to provide a true sense of volumetric depth. An effective user interface, however, has yet to be thoroughly studied. Tangible user interfaces (TUIs) enable a tactile interaction with a hologram through an object. The object has physical properties indicating how it might be used with multiple degrees-of-freedom. We propose a TUI using a planar object (PO) for the holographic medical volume visualization and exploration. We refer to it as mxdR hologram slicer (mxdR-HS). Users can slice the hologram to examine particular regions of interest (ROIs) and intermix complementary data and annotations. The mxdR-HS introduces a novel real-time ad-hoc marker-less PO tracking method that works with any PO where corners are visible. The aim of mxdR-HS is to maintain minimum computational latency while preserving practical tracking accuracy to enable seamless TUI integration in the commercial mxdR-HMD, which has limited computational resources. We implemented the mxdR-HS on a commercial Microsoft HoloLens with a built-in depth camera. Our experimental results showed our mxdR-HS had a superior computational latency but marginally lower tracking accuracy than two marker-based tracking methods and resulted in enhanced computational latency and tracking accuracy than 10 marker-less tracking methods. Our mxdR-HS, in a medical environment, can be suggested as a visual guide to display complex volumetric medical imaging data.

## 1 Introduction

Mixed reality (mxdR) creates a 'mixed environment' where the virtual and user's real-world environments are merged by seamlessly intermixing virtual contents in the user's physical surroundings [1]. Advances in mxdR technology with the introduction of affordable commercial mxdR head-mounted displays (mxdR-HMDs, e.g., Microsoft HoloLens [2] and Magic Leap One [3]) now allow the generation of holographic images (holograms) in the real-world environment with a sense of native three-dimensional (3D) depth. mxdR technology is not being widely adopted in the medical environment to display volumetric medical imaging data with holograms for potential image interpretation and in medical education [4,5]. In these applications, data visualization and exploration typically involve volume navigation, selection of regions of interest (ROIs), adding labels and annotations to the ROIs, and augmenting pre-surgical planning information.

The standard user interfaces (UIs) for mxdR-HMDs include hand gestures, eye gaze, and voice commands [6,7]. These UIs lack physical interaction, which can provide additional sensory signals to the user, with the hologram [8]. Handheld objects (e.g., a paddle [9], pen [10], etc.) have been used to provide the sensory input for the physical interaction with virtual content and these are referred to as tangible user interfaces (TUIs) [11]. A TUI tracks the object and registers a virtual interaction tool to it. The TUI provides a form factor with physical properties that indicate how it can be used for the interaction [7]. Several investigators have shown that a TUI is effective, and it allows simultaneous control of multiple degrees-of-freedom (DoF) [12-16]. Bach *et al.* [17]

reported that a TUI improves the time and accuracy of interacting with the hologram when coordination between 3D perception and high DoF interaction is required. We propose using a rectangular object, as the planar object (PO), to freely interact with volumetric medical imaging data. The PO is flat, and its contours are similar to the 'cutting plane' that is the default tool for desktop-based medical imaging software which uses axis-aligned crosshairs manipulated with a computer mouse [18].

The main technological challenge for enabling a TUI on mxdR-HMDs, which inherently have limited computational resources, is tracking the PO timely and accurately while considering the trade-off between computational latency (or lag) and tracking accuracy to minimize misregistration [17,19-21]. The most widely used method for tracking the PO is a marker-based tracking method that attaches a binary fiducial marker, such as a Quick Response (QR) code [22], to the PO [23,24]. The marker is then tracked using an image feed from a camera. The pose (position and rotation) and geometry of the PO are estimated based on the prior knowledge of marker size and geometrical relation between the marker and the PO [23]. The marker-based tracking method is the default baseline solution for its ability to generally satisfy both computational latency and tracking accuracy. However, its main drawbacks are that the placement and accurate geometrical alignment of the marker needs to be done individually for every PO before use. Marker-less tracking methods that directly track the PO have been developed where a model of the PO is pre-defined and used for recognizing and tracking the PO from the image feed [23,25]. Although the marker-less tracking methods improved usability, they require more intensive computation and generally produce inferior tracking accuracy when compared to the marker-based methods, and thus restricting their usage to tracking the stationary POs [26-29].

Medical imaging is now indispensable to modern healthcare and is essential for effective patient management [30]. Medical imaging modalities now produce vast amounts of volumetric data as the imaging devices become more sophisticated and often combine two imaging modalities such as with PET-CT where positron emission tomography (PET) is coupled to computed tomography (CT), and similarly with PET-MR which has PET combined with magnetic resonance (MR) imaging [31]. 3D visualization tools are increasingly being included in the vendor software with these scanners, however, the use of mxdR-HMDs and the generation of 'interactive medical holograms' is new and not considered a routine component of an imaging specialists image interpretation environment [4]. These medical holographic displays are based on a standard UI consisting of hand gestures and voice commands [32,33]. Our contention is that holographic displays can play an important role in the visualization and perhaps the interpretation of complex medical imaging data if there are effective user interfaces.

In this study, we propose an mxdR hologram slicer (mxdR-HS), comprising of a new marker-less PO-based TUI that allows interactions with the holographic volume visualization. The mxdR-HS works with any PO where corners are visible and does not rely on a PO's 3D model as a priori. The aim of mxdR-HS is to maintain minimum computational latency while preserving practical tracking accuracy to enable seamless TUI integration in the commercial mxdR-HMD, which has limited computational resources. This aim is drawn from the effect of latency on TUI integration where computational latency can result in the virtual interaction tool to not correctly align to the PO in motion, and thus causing the virtual tool to appear unstable, jittering or 'swimming around' [17,34,35]. Further, user interaction performance was found to be linearly decaying with increasing latency [36]. The latency also has significant effects on the use experience, such as perceived causality when the virtual interaction tool is moved by the PO [20].

We implemented our mxdR-HS on a commercial mxdR-HMD, Microsoft HoloLens, with its built-in depth camera. Our mxdR-HS was evaluated for computational latency and tracking accuracy. In summary, the main contributions of this study are:

- Propose a depth-based real-time ad-hoc marker-less PO tracking method to enable the seamless integration of TUI on the commercial mxdR-HMD.
- Introduce lightweight (computationally efficient) image processing techniques to find PO's corners to enable timely PO tracking and pose estimation with affordable tracking accuracy in the commercial mxdR-HMD.
- Comparative evaluation of our mxdR-HS to 10 marker-less [37-46] and 2 marker-based [47] tracking methods. We demonstrate that our method outperformed all marker-less methods by large margins and that it improved computational latency but with marginally lower tracking accuracy when compared to marker-based counterpart.
- Demonstrate the usage of our method in use cases of visualization of volumetric medical imaging data and multimedia data.

## 2    Related work

Several investigators have reported using handheld objects as TUIs and these approaches have been reviewed by Zhou et al. [7]. AlSada and Nakagima [48] used a PO, as a web browser by projecting web pages onto it. Yamada et al. [49] used a PO as a virtual panel, visualizing the information related to a product or place nearby it. Bach et al. [17] explored the effectiveness of a PO to explore the 3D scatterplot data in an mxdR environment, where interactions included ROI pointing, clustering, selecting, and cutting. The main aim of these studies was to explore using a PO as a TUI for mxdR-MHD, and so marker-based tracking methods were used for the PO tracking.

Depth cameras measure the distance between the camera and objects and thus produce a depth image. The depth image is a greyscale (numeric) image where each pixel (px) contains the depth while a red, green, and blue (RGB) image contains colors. The depth image can be used to identify the PO directly based on its geometry [50]. The depth image is converted to a set of discrete 3D points, a point cloud, that represents the geometry of the real-world environment in which the PO is blended with the other objects [51]. The PO can then be recognized and tracked from the point cloud. Iterative closest point (ICP) and its variants [37-41] are commonly used to identify and track the PO [23]. In a typical ICP, a 3D model of the PO must be provided, and the ICP then iteratively searches for the PO in the point cloud by estimating the minimal differences between the point cloud and the model [37]. The main drawbacks of ICP, however, are its need for a predefined PO 3D model and its dependency on the initial alignment between the model and the PO in the point cloud [52]. It is challenging to obtain a good initial alignment when the PO and the mxdR-HMD are constantly in motion. Hence, ICP is generally employed where an object is in a fixed position [27,53]. An alternative method for marker-less tracking is a mathematical model of primary shapes (e.g., a sphere, a plane, etc.). The PO is detected by randomly selecting points from the point cloud as a subset to reconstruct a shape primitive and match it with a mathematical model of the PO. An example of a mathematical model for mxdR-HMD is the random sample consensus (RANSAC) method [44] which enables coarse estimation of a PO that is a dominant object in a fixed position, such as with walls [28] and table [28,29].

# 3 Methods

We implemented the mxdR-HS prototype on a Microsoft HoloLens because of access to the built-in depth camera via support for the application programming interface (API). We note that our mxdR-HS can be implemented to other mxdR-HMDs that have a built-in depth camera. HoloLens's depth camera operates in a short-throw mode (up to 1 m fields of view) for hand tracking and a long-throw mode (up to 3.5 m) for environmental sensing [54]. We used the short-throw mode exclusively for PO tracking given the proximity of the handheld PO to the camera, the higher capture frame rate, and reduced noise characteristics when compared to the long-throw mode [54]. Although employment of other built-in cameras (e.g., greyscale stereo cameras) may enhance the tracking accuracy of our mxdR-HS, it solely used the depth camera to avoid the HoloLens's system instability reported by Liebmann *et al.* [55] and preserve the minimum computational latency by avoiding the additional camera synchronization process reported by Labini *et al.* [56].

## 3.1 The real-time ad-hoc marker-less PO tracking method

The mxdR-HS's real-time ad-hoc marker-less PO tracking method has 3 modules as outlined in Fig. 1. There is an Image streamer, a Tracker engine, and a Content renderer. The Image streamer obtains the depth image and a transformation matrix that map a 3D point in the depth camera (*D. cam*) to the real-world environment (*World*). The Tracker engine receives the depth image from the Image streamer. It identifies the corners of the PO from the depth image and then samples each corner's pixel (px) value (depth). The Content renderer receives the location and depth of the corners of the PO from the Tracker engine and the transformation matrix from the Image streamer. It transforms the corner location from 2D points in the depth image to 3D points in World and augments the virtual interaction tool to the PO.

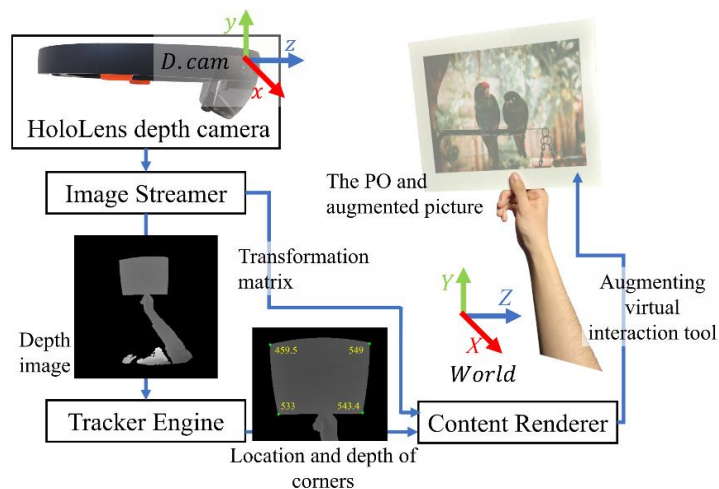

**Fig. 1** The mxdR-HS's real-time ad-hoc marker-less PO tracking method; a picture (of 2 birds [57]) is augmented to the PO

### 3.1.1 Image streamer

In the Image streamer, the obtained depth image has a 488×450 px resolution with each px containing an intensity value (an integer between 0-1,000). The intensity values represent the depth on a millimeter (mm) scale. The transformation matrix is a 4×4 matrix obtained by calling HoloLens's API via *HoloLensForCV* library [58].

### 3.1.2 Tracker engine

The Tracker engine tracks the PO corners from the depth image and has 4 steps with intermediate results as shown in Fig. 2. Step 1 is depth image trimming, step 2 is PO candidate identification, step 3 is PO corner detection, and step 4 is patch-based intensity enhancement. The Tracker Engine uses the *OpenCVSharp-UWP* library [59] to exploit the library's support of various computer vision algorithms in mxdR-HMDs.

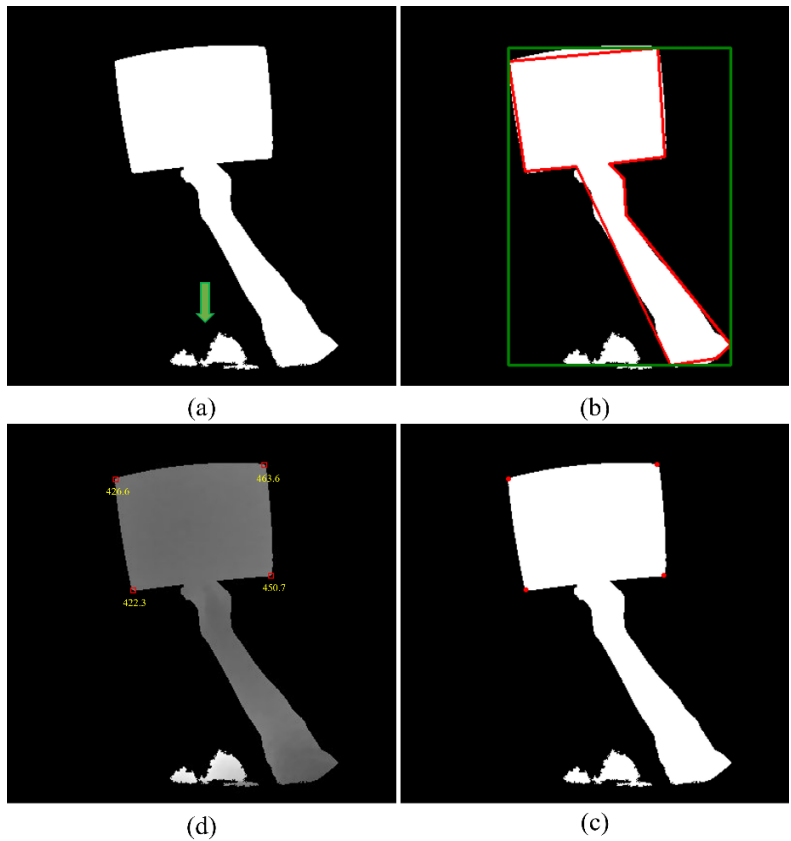

**Fig. 2** Silhouette of the depth image processing steps, in a clockwise direction of the Tracker Engine showing a silhouette of the PO, the upper limb (hand, forearm, and arm) holding the PO and user's abdomen entering the field-of-view (FoV) at the bottle of the image

*Depth image trimming*: The depth image has a padding area filled with zero intensity values, and this area was trimmed to reduce future computational costs. The trimmed image was filtered with a threshold set to the intensity values of < 150 and > 1000. The thresholding range was empirically set to remove invalid intensity values for objects out of the depth camera's field-of-view (FoV) and noise, and to keep valid intensity values that belong to the PO. After thresholding, the depth image was converted to a binary mask as shown in Fig. 2a.

*PO candidate identification*: For each connected region in the binary mask, an outline was generated using Suzuki and Abe's second border following algorithm [60] and its size was calculated in px. Small regions (< 3,500 px size) were filtered out, for example, the user's abdomen at the bottom of the binary mask (a green arrow in Fig. 2a). The outline of the remaining regions was straightened (a red contour in Fig. 2b), based on the Douglas-Peucker algorithm [61], to remove the curves that impede the reconstruction of the square/rectangular PO. We considered algorithms that directly search for PO's geometry, such as using the Hough transform as in Jung and Schramm [62] or fitting a plane similar to Hettiarachchi and Wigdor [29]. However, due to the hand's hindrance to the geometry search and the difficulties of separating the hand from the PO where the boundary is ambiguous in the depth image, we adopted our simple yet effective border following algorithm. Further, the depth image contains varied errors over the images which prevent the precise fitting of the plane to the PO (see section 4.3 and 5.3 for details). Hence, an axis-aligned bounding rectangle (a green box in Fig. 2b) was generated on each straightened outline to identify the region containing the PO and the User's upper limb by checking the location of the bounding rectangle's bottom corners. The region was identified as a PO candidate if the bottom corner(s) of the bounding rectangle is located near the bottom corners of the binary mask where the upper limb enters the FoV.

*PO corners detection*: Each simplified outline of the PO candidates has vertices, and we implemented a method for detecting PO corners by examining each vertex based on its height, internal angles, and sequences. Firstly, the vertex was filtered out if it was located below one-fifth height of the bounding rectangle. This partially removed the vertices on the upper limb. Secondly, the internal angle of each remaining vertex was calculated. The vertex was flagged as a potential corner if its internal angle was between 30° and 150°. We added ±60° to 90° as a buffer to prevent a false-negative flagging. Thirdly, each potential corner was checked to determine if its predecessor or successor vertex was a potential corner. If four sequential potential corners were obtained from the predecessor/successor check, they were considered as final corner candidates such that the PO corners were located next to each other in the simplified outline. If this did not occur, each potential corner and its 4 successor vertices were checked to determine if there were 4 potential corners.

After all the simplified outlines were processed, a centroid of the simplified outline was obtained if there were groups of final corner candidates. A group with the highest centroid relative to the bounding rectangle became the group of PO corners. This is because the PO is presented at the top of the region and the others are the hand which is smaller than the PO. The PO corners were fed into Forstner's corner refinement algorithm [63] to iteratively refine corner locations to the near intersections of edge elements (red points in Fig. 2c).

*Patch-based intensity enhancement*: The patch-based intensity enhancement stage sampled the intensity value of each PO corner with a square patch using the corner as its center (red boxes in Fig. 2d). It was designed to mitigate the low quality of the depth camera where the intensity values of the PO were randomly changing over time even though the PO and depth camera were stationary (more details in section 4.3). Within the patch, only non-zero intensity values were summed, averaged, and used as the corner's intensity value (yellow texts in Fig. 2d). We used a patch size of 5×5 px, empirically derived, to be the best balance between computational latency and tracking accuracy; 7 size variations from 1×1 to 13×13 px with an increment of 2 px were experimented; these results are included in Appendix A.

### 3.1.3 Content renderer

The Content renderer had three steps to process the transformation and was based on the pinhole camera model. In step 1, each corner in the depth image was transformed to an undistorted point on the image plane, $P_{Image\ plane}$ ($U_i$, $V_i$, -1) using the *HoloLensForCV* library [58]. The image plane is a plane where the PO is projected through the depth camera's aperture and digitized to the depth image. In step 2, $P_{Image\ plane}$ was unprojected to a 3D point in *D. cam*, $P_{D.cam}$, through the following calculation:

$$Corner\ intensity \times \frac{1}{\sqrt{(U_i)^2+(V_i)^2+(-1)^2}} \times -P_{Image\ plane} \qquad (1)$$

where the corner intensity is the distance between the depth camera's aperture and $P_{D.cam}$. In step 3, a 3D point in *World*, $P_{World}$, was obtained through the following calculation:

$$\textbf{The transformation matrix} \cdot P_{D.cam} \qquad (2)$$

The pose and size of the PO were calculated based on $P_{World}$. The Content renderer rendered a hologram of a virtual interaction tool and augmented it to the PO. Altogether, example applications of mxdR-HS for medical volume visualization and multimedia viewer were implemented and shown in section 6.

## 4 Experiments

We used two polystyrene POs: a rectangle with dimensions 300×240 mm (PO#1, Fig. 3a) and a square 220×220 mm (PO#2, Fig. 3b). We carried out three experiments: i) We measured the tracking performance of our mxdR-HS compared to the marker-based tracking methods implemented with *ArUco* [47]. ii) We tested the feasibility of 10 marker-less tracking methods using the depth image [37-46] in PO tracking on the mxdR-HMD. iii) We measured the changes in the intensity values of PO while the PO and depth camera were in a fixed position.

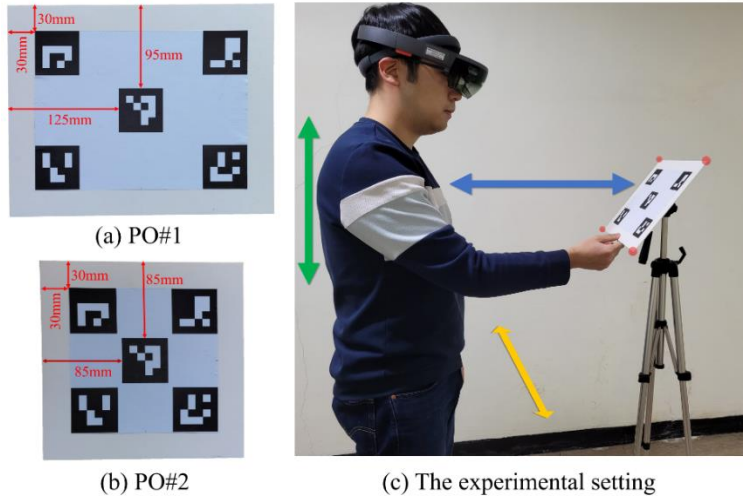

**Fig. 3** Overview of POs and the experimental setting showing the user holding the POs in the right hand with the arm and forearm outstretched, the two POs and the user's movements. The lower anterior abdominal wall of the user is seen below the right elbow

### 4.1 mxdR-HS and marker-based tracking methods experiment

We measured mxdR-HS's computational latency and tracking accuracy with PO#1 and PO#2. We defined 'computational latency' as the time elapsed from the commencement of the Tracker engine to obtaining the $P_{World}$ in the Content renderer. We measured tracking accuracy by calculating the Euclidean distance between the $P_{World}$ localized by mxdR-HS and the manually localized corners which are assigned as the ground truth.

We selected the *ArUco* implementation [47] as the marker-based tracking method because, based on published data [47,64], it had better environmental adaptation and tracking performance when compared to the other alternatives - *ArToolKit+* and *AprilTags* [65,66]. The experimental software was implemented and deployed to the HoloLens using *Unity 2018.4.30f* with *MRTK 2.5.3* [67,68]. It captured the depth and color images concurrently. The depth images were used by mxdR-HS while the RGB images were used by the marker-based tracking method.

  For our first experiment (i), PO#1 and PO#2 had five *ArUco* markers (50×50 mm), one center marker and four corner markers with a margin of 30×30 mm to the edges. Fig. 3 shows: (a) PO#1 and (b) PO#2 with the center, corner markers and the margins from the edges, and (c) experimental setting where the blue, yellow and green double arrows indicate the user's random movements and red spheres on corners of the PO indicating the ground truth. Two marker-based tracking methods, center marker-based and corner marker-based, were measured. During the center marker-based tracking, only the center marker was used, and for corner marker-based tracking, the corner markers were used. We attached PO#1 or PO#2 to a tripod one at a time.

  The HoloLens' display was calibrated to the user's eyes using the HoloLens' default software. Then, the user manually localized the PO corners to the HoloLens by placing four holographic spheres with a diameter of 5mm (red spheres in Fig. 3c) on each corner. The user then walked around with each PO at various angles to confirm that the spheres were accurately aligned, and to correct the sphere placements, as necessary. The user gently grabbed the PO to replicate the usage scenario such that holding the PO is captured in the depth image. The user varied the distances and viewing angles between the cameras and the PO by randomly moving around the PO while gazing at its center. The depth and color images were captured every 0.5 seconds, and computational latency and tracking accuracy were measured. The sequence was completed when 100 image pairs of depth and color images (a total of 200 images) were obtained. There were 4 sub-experiments using mxdR-HS with each of the two marker-based tracking methods measured for each of the two POs. We repeated the sequence three times for each sub-experiment to mitigate possible corner localization errors caused by the manual placement of the virtual spheres; 12 sequences were carried out.

  In total, we captured 1,200 image pairs. Among the pairs, 70 pairs had one of the following errors: a failure of marker-based tracking methods (6 pairs), a malfunction of the RGB camera (7), an out-of-view marker (2), delivery of corrupted depth image from HoloLens (50) and a failure of intensity value sampling with the 1×1 px patch size (i.e., the intensity value was zero) (5). The mxdR-HS & center maker-based tracking method sub-experiments produced 29 errors, and their image pairs without the errors were used for measuring computational latency and tracking accuracy of the mxdR-HS and the center marker-based tracking method. The mxdR-HS & corner maker-based tracking method sub-experiments produced 41 errors, and their image pairs without the errors were separately used for measuring computational latency and tracking accuracy of the corner marker-based tracking method. The mxdR-HS TUI interaction was designed to allow the user to freely translate and rotate the PO while holding it. However, we reversed the interaction during the experiment such that the PO was fixed, and the user moved around it. This reversal enabled the measurement of the performance of the PO tracking in stable conditions, i.e., to ensure that a consistent ground truth position was used in all the sub-experiments.

## 4.2 Marker-less tracking methods experiment

For our second experiment (ii), the performance of the 10 marker-less tracking methods using the depth image was measured with the same set of depth images used for measuring the mxdR-HS performance. The methods were selected based on their use in published reports [27-29,53] and accessibility of the implementation for precise reproduction. They were grouped into two groups. The first was a 3d model group with 7 methods that relied on a 3D model of the PO, including point-to-point ICP [37], point-to-plane ICP [38], ICP with normal [39], generalized-ICP [40], Go-ICP [41], Normal Distributions Transform (NDT) [42] and Coherent Point Drift (CPD) [43]. The second was a mathematical model group with 3 methods that used a mathematical model of the plane, including RANSAC-based plane segmentation [44], Efficient RANSAC [45], and Global-L0 [46]. These methods required the depth images to be converted to point clouds. We measured the computational latency for tracking the POs with the 10 marker-less tracking methods and the depth image to point cloud conversion, separately. The tracking accuracy was calculated differently based on the group. For the 3D model group, tracking accuracy was defined as the Euclidean distance between the manually localized corner and the 3D model's corner. For the mathematical model group, tracking accuracy was defined as the Sorensen–Dice coefficient (Dice) [69] between the PO in the point cloud and the resultant segmentation; and calculated by the following formula:

$$\frac{2|PO \cap segmentation|}{|PO| + |segmentation|} \quad (3)$$

For comparison with mxdR-HS, the Dice was calculated by segmenting the PO based on the corners localized by mxdR-HS.

The 10 marker-less tracking methods and depth image to point cloud conversion were implemented on a laptop computer (Intel i5-6300HQ with 16GB RAM) and not in the HoloLens. This is because the implementation of these methods was optimized for the PC computing environment and there was no implementation available in the HoloLens. The *point cloud library* [70] was used for carrying out the point-to-point ICP, point-to-plane ICP, ICP with normal, generalized-ICP, NDT, and RANSAC-based plane segmentation methods. The rest of the methods, Go-ICP, CPD, Efficient RANSAC, and Global-L0, were executed based on public code released on the author's webpage [71] and GitHub repositories [72-74]. The depth image to point cloud conversion was implemented with the Cython programming language. The 3D models of PO#1 and PO#2 were generated as a polygon mesh using *Autodesk Meshmixer* [75]. The polygon mesh was converted to a point cloud by sampling 11,588 points for PO#1 and 9,670 points for PO#2. The number of sampling points was derived from randomly selected 10 depth images taken while the POs were in motion as the mxdR-HS's interaction design was originally intended. The initial location of 3D models was set based on the average location of POs in the selected depth images.

## 4.3 Random error measurement experiment

The tracking accuracy of mxdR-HS relied on how much the corner's intensity value was accurately measured by the depth camera as it used the corner's intensity value for localizing the corner in 3D. The volatile change of intensity value is caused by a random factor (e.g., environmental lights, air temperature, and humidity) interfering with the depth camera's depth measurement [50]. Therefore, we formulated an experiment to measure the changes based on the recommendation from Microsoft's documentation [76] for our third experiment (iii). The Random error is the standard deviation of the intensity values in a sample point through a series of depth images and calculated through the following formula:

$$\text{Random error} = \sqrt{\frac{\sum_{t=1}^{N}(d_t - \bar{d})^2}{N}} \quad (4)$$

where *N* denotes the number of depth images in the series, $d_t$ is the intensity value of each pixel compositing a depth image $t$ and $\bar{d}$ is the average intensity value of $d_t$ through the series of depth images. A series of 600 depth images were captured after setting HoloLens and PO#1 to be 500 mm apart, which is the middle of the maximum range of the depth camera. The HoloLens and PO#1 were in a fixed location where the HoloLens was placed 142 mm higher than PO#1 and facing approximately perpendicular to PO#1. PO#1 was manually segmented from the first depth image. The segmentation was used to calculate the Random error of PO#1 through the series.

# 5  Results

## 5.1  Result of mxdR-HS and marker-based tracking methods experiment

The mxdR-HS's average computational latency (lower is better) for PO#1 and PO#2 is plotted in Fig. 4 with center and corner marker-based tracking methods. The mxdR-HS was faster than the center marker-based tracking method by 34.7 ms for PO#1 and 35.4 ms for PO#2; and the corner marker-based tracking method by 37.5 ms for PO#1 and 39.6 ms for PO#2. The mxdR-HS had the best standard deviation, minimum and maximum computational latencies.

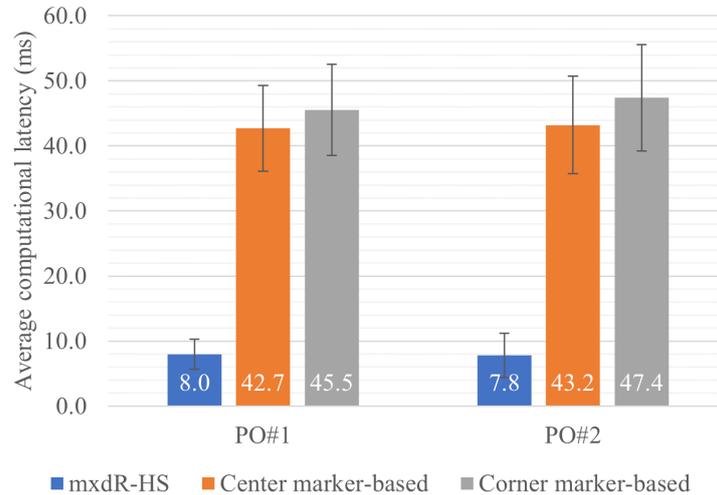

**Fig. 4**   Average computational latency of mxdR-HS for PO#1 and PO#2 over the center and corner marker-based methods with standard deviation as error bars

The average tracking accuracy (lower is better) of mxdR-HS is plotted in Fig. 5 with center and corner marker-based tracking methods. The mxdR-HS had poorer average tracking accuracy when compared to the center marker-based tracking method by 3.1 mm for PO#1 and 4.7 mm for PO#2; and the corner marker-based tracking method by 3.9 mm for PO#1 and 5.6 mm for PO#2. The mxdR-HS had the best standard deviation and maximum tracking accuracies for PO#1 and PO#2. The best minimum tracking accuracy was found from the corner marker-based tracking method.

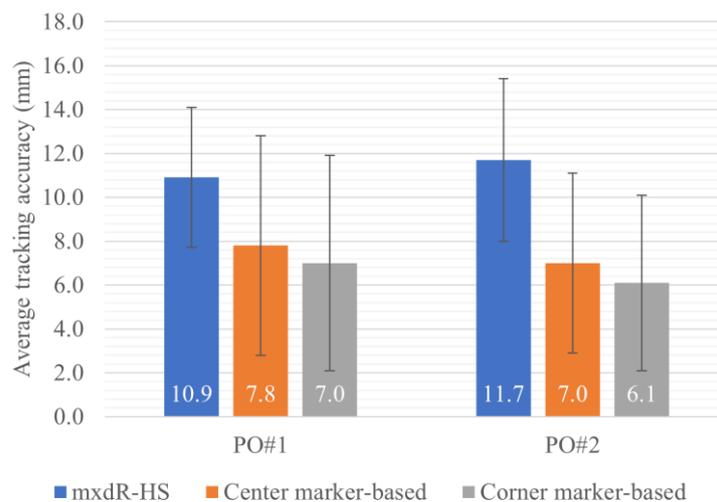

**Fig. 5**   The average tracking accuracy of mxdR-HS for PO#1 and PO#2 over the center and corner marker-based tracking methods with the standard deviation as error bars

## 5.2 Result of marker-less tracking methods experiment

The computational latencies of mxdR-HS were compared with the 10 marker-less tracking methods and these results are shown in Table 1. For PO#1 and PO#2, the RANSAC-based plane segmentation method had the best average and the standard deviation, with mxdR-HS being second best. However, it is important to note that the 10 marker-less tracking methods, including the RANSAC-based plane segmentation method, were not computed using a HoloLens due to their availability of implementation on the HoloLens and instead computed using a laptop which has vastly superior computational resources than the HoloLens. The compute power scores from *Unity's benchmark tool* [77] showed the laptop had approximately 8.1 times higher compute power than the HoloLens. The 10 marker-less tracking methods all required an additional conversion procedure between the depth image and point cloud, which took 649.0 ms (SD=46.5 ms) for PO#1 and 612.1 ms (SD=31.2 ms) for PO#2. This conversion time was excluded from the computational latency.

**Table 1** Computational latency of mxdR-HS, executed in the HoloLens, and the 10 marker-less tracking methods, executed in the laptop, in milliseconds (lower is better)

| Tracking method | PO#1 Average | PO#1 Standard deviation | PO#2 Average | PO#2 Standard deviation |
|---|---|---|---|---|
| RANSAC-based plane segmentation | **4.8** | **1.4** | **5.1** | **0.9** |
| mxdR-HS[1] | 8.0 | 2.3 | 7.8 | 3.4 |
| Efficient RANSAC | 120.4 | 24.1 | 112.0 | 18.2 |
| NDT | 282.0 | 27.3 | 241.6 | 32.2 |
| Point-to-point ICP | 435.6 | 275.7 | 294.1 | 63.3 |
| Global-L0 | 532.1 | 126.7 | 477.8 | 90.8 |
| Generalized ICP | 679.7 | 293.0 | 485.9 | 106.6 |
| ICP with normal | 779.1 | 539.1 | 467.8 | 323.5 |
| Go-ICP | 21179.4 | 211.2 | 21188.0 | 473.8 |
| Point-to-plane ICP | 38832.1 | 21192.1 | 37105.4 | 21102.0 |
| CPD | 99995.0 | 20547.2 | 75132.0 | 12500.7 |

Note: 3D model group results are shown in green; mathematical model results in blue; 10 marker-less tracking methods listed in ascending order of the average computational latency for PO#1.
[1] Our mxdR-HS was executed in the HoloLens whose compute power scores [77] was 12.3% of the laptop.

The tracking accuracies of the mxdR-HS and the marker-less tracking methods are shown in Table 2. For the 3D model group, Go-ICP had the best average tracking accuracy for PO#1 and PO#2, and for the mathematical model group, Global-L0 had the best average tracking accuracy. The mxdR-HS's average tracking accuracy was better than Go-ICP's by 15.4 mm for PO#1 and 35.1 mm for PO#2; and Global-L0's by 0.031 Dice for PO#1 and 0.029 Dice for PO#2.

**Table 2** Tracking accuracy of mxdR-HS and the 10 marker-less tracking methods

| Tracking method | PO#1 Average | PO#1 Standard deviation | PO#2 Average | PO#2 Standard deviation |
|---|---|---|---|---|
| mxdR-HS | **10.9** | **3.2** | **11.7** | **3.7** |
| Go-ICP | 26.3 | 23.3 | 46.8 | 41.7 |
| Point-to-plane ICP | 35.1 | 21.1 | 143.5 | 95.6 |
| Point-to-point ICP | 62.0 | 15.0 | 134.9 | 40.6 |
| Generalized ICP | 66.4 | 20.1 | 138.5 | 39.3 |
| NDT | 134.7 | 37.7 | 153.1 | 39.0 |
| CPD | 1059.6 | 337.4 | 1319.5 | 237.1 |
| ICP with normal | 1986.2 | 32659.0 | 3483.2 | 55313.5 |
| mxdR-HS | **0.985** | **0.008** | **0.984** | **0.008** |
| Global-L0 | 0.954 | 0.043 | 0.955 | 0.044 |
| RANSAC-based plane segmentation | 0.944 | 0.158 | 0.660 | 0.449 |
| Efficient RANSAC | 0.851 | 0.031 | 0.825 | 0.030 |

Note: 3D model group (in green labels) presented in millimeters (lower is better); mathematical model group (in blue) is presented in Dice (higher is better); 3D model group was sorted in ascending order of average tracking accuracy for PO#1 whereas mathematical model group was sorted in descending order.

## 5.3 Results of Random error measurement experiment

The Random error (lower is better) of the PO#1 is calculated and visualized in a grayscale image as shown in Fig. 6. The average Random error of the PO#1 was 1.8 mm (SD=1.0 mm) with the error ranging between 1.2 to 34.9 mm. The boundary including the corners had higher Random error than the rest of PO#1 except where the center marker was attached, as shown in Fig. 6.

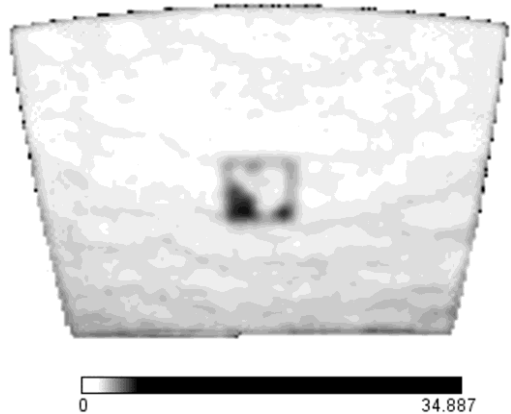

**Fig. 6**   The Random error of PO#1. The darker gray color indicates the greater Random error bar

# 6   Example applications of mxdR-HS

## 6.1   3D holographic visualization of volumetric medical imaging data

We used a tablet PC (Dell Venue 11 Pro) as the PO; it acted as the volume slicing tool and information augmentation interface. We designed this approach to resemble a common visualization interaction that occurs when visualizing volumetric medical images for interpretation. A hologram of the head and neck and upper torso of a PET-CT dataset of a patient with lung cancer is shown in Fig. 7.

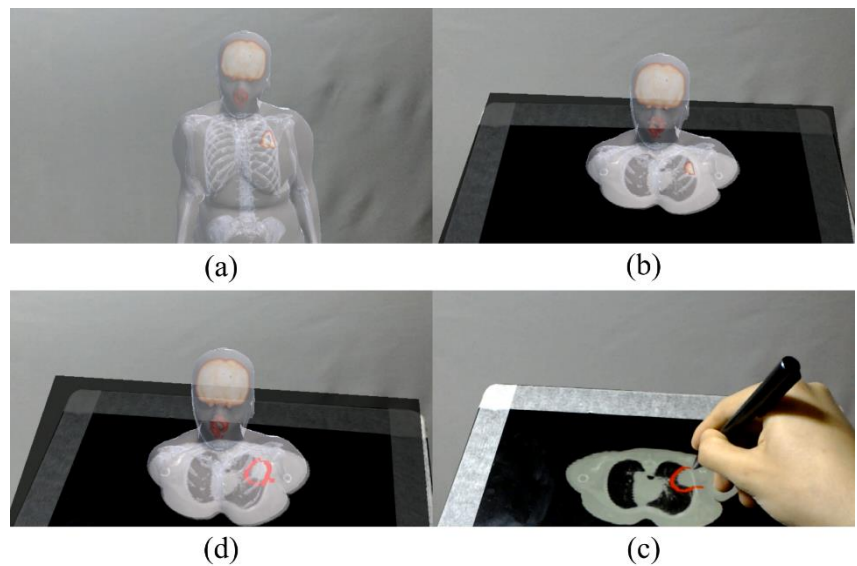

**Fig. 7**   The mxdR-HS in action showing the 3D visualizations and the 2D image for placement of the ROI

In Fig. 7, the components of the visualization are depicted. In a), there is a view through the HoloLens showing the 3D hologram of the head and neck / upper torso and a tumor in the left lung (small red spot). The PET was rendered using the maximum intensity projection (MIP) rendering technique and CT with a surface rendering technique of bone and skin structures. The two renderings were then fused into one. In b), the tablet PC (as the PO) was used to slice the hologram and the 2D CT image slice is augmented to the tablet as complementary data. The user can adjust the volume slicing tool with the tablet PC where it can be freely positioned. In c), a 2D CT image slice of the thorax showing the lung tumor is illustrated with the user drawing a red ROI on the tumor. In d), the annotation of the lung tumor is superimposed on the hologram. The masking tape on the edge of the tablet PC was used to prevent the depth loss caused by the glossy surface of the tablet's screen. A demo video is accompanied in the online resource 1.

## 6.2 Multimedia viewer

We used a picture showing two parrots with a backdrop of trees scene as an example of our method being used to depict multimedia content. The picture was obtained from a public repository with a public domain license [57]. The picture was rendered to a PO (Fig. 8a), and it was translated and rotated according to the pose of the PO when the user moved it. Without the PO, the picture was rendered in mid-air and the background trees were visible through the hologram (Fig. 8b).

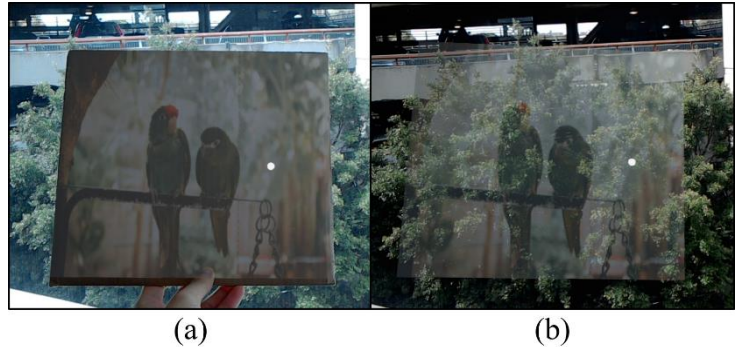

(a)                  (b)

**Fig. 8** A picture hologram is displayed on the PO in a) and in mid-air in b)

## 7 Discussion and future work

Our main findings are that our mxdR-HS: (i) tracked the marker-less POs at an interactive rate on a commercial mxdR-HMD; (ii) achieved marker-less tracking of 10 mm-level tracking accuracy; (iii) outperformed the computational latency but was marginally less accurate over the marker-based tracking methods; (iv) showed enhanced computational latency and tracking accuracy than the 10 marker-less tracking methods, and (v) could be applied to interact with and explore complex medical imaging data and multimedia content.

Our mxdR-HS provided real-time marker-less PO tracking (8.0 ms on PO#1 and 7.8 ms on PO#2 in average computation latency), which we suggest is suitable for use with an mxdR-HMD and did not exceed the computational latency limits of 30 – 40 ms that Pintaric and Kaufmann [78] indicated was the threshold to ensure that the user's experience was positive. It was able to update the location of the PO faster than the recommended hologram refresh rate of the HoloLens, 60 FPS ($\approx$ 16.6 ms) [79], and thus enabled seamless TUI integration [17,20]. The computational latency of the mxdR-HS was also superior to the marker-based tracking methods in meaningful difference to the user ($\geq$ 34.7 ms on PO#1 and $\geq$ 35.4 ms on PO#2). Mania *et al.* [80] reported that humans can perceive variations in latency as small as 15 ms and Jota *et al.* [81] found that people could perceive latency as low as 2.4 ms when dragging the virtual content on the physical surface. We suggest that the superiority of our method relates to its optimization for localizing the PO corners with the depth images.

Our method had more consistent tracking accuracy than the marker-based tracking methods (lower standard deviation) but had a poorer average tracking accuracy by $\leq$ 3.9 mm on PO#1 and $\leq$ 5.6 mm on PO#2. The performance of average tracking accuracy from our mxdR-HS is related to, in part, the low-quality depth images obtained from HoloLens's built-in depth camera as shown in the Random error (see Fig. 6). This same issue with low-quality depth images is consistent with the findings of Gu *et al.* [27] and Gsaxner *et al.* [53]. Holloway *et al.* [82] reported a trade-off effect between computational latency and tracking accuracy; a rule of thumb is that 1 ms of latency introduces 1 mm of accuracy reduction in the worst case. The estimated trade-off comparison suggests that our mxdR-HS may be comparable or even better than the marker-based tracking methods. Our mxdR-HS had a better trade-off by 31.6 and 33.6 mm for PO#1; and 30.7 and 34.0 mm for PO#2 than the center marker-based and corner marker-based, respectively.

We found that the mxdR-HS had better computational latency and tracking accuracy than the 10 marker-less tracking methods using depth images. Although the RANSAC-based plane segmentation method [44] showed better computational latency, this was due to it being computed using greater computing power. When we used *Unity's benchmark tool* to simulate the RANSAC-based plane segmentation method to use the same compute power available from HoloLens, we estimated its computational latency would be 38.9 ms on PO#1 and 41.3 ms on PO#2. Hence, our mxdR-HS may have better computational latency by 30.9 ms on PO#1 and 33.5 ms on PO#2. This is almost 4 folds over our method. Further, the computational latency of depth image to point cloud conversion should add additional computational latency to the 10 marker-less tracking methods as it required the point cloud as an input. This additional computational latency would be 2-3 seconds on the HoloLens as reported by Kastner *et al.* [83]. Therefore, the direct implementation of 10 marker-less tracking methods on the HoloLens would be impractical for achieving real-time PO tracking.

The current tracking accuracy of mxdR-HS means that it should not be used in the medical context of planning precise medical interventions (e.g., radiotherapy) and interventional procedures (e.g., biopsies). Nevertheless, the current accuracy does not negate its applications for providing clinicians a valuable holographic 3D display of an ROI. In the 3D holographic medical visualization application, the PO provided a physical interaction with the hologram as the slicing tool. There was also a physical platform that the user employed to make annotations. We suggest that these findings make our mxdR-HS a useful adjunct for the visualization of complex imaging data. In the multimedia viewer example, the PO provided an opaque background and could block the background elements behind the multimedia content. This attribute could be used to improve the visibility of the hologram where the visual interference from the background elements is inevitable due to the hologram's transparency. However, the practicality and the usefulness of the mxdR-HS were not thoroughly measured as part of the scope of this work. We plan to investigate more about how mxdR-HS is suited for interacting with the holograms, including the volume slicing technique, through a user study in future work.

    We suggest that the application of neural network (NN)-based tracking methods on our mxdR-HS may improve the tracking accuracy. NN-based tracking methods have been employed in mxdR-HMDs, including the HoloLens, to enable accurate tracking of complicated objects, such as face tracking for augmenting 3D imaging of the patient's head [53] and robot tracking for calibrating the mxdR-HMD to the real-world environment [83]. However, these NN-based tracking methods are computationally intensive and their direct implementation on mxdR-HMDs would be impractical due to their incapability to achieve the minimum computational latency for our TUI usage [53,83-87]. Further, their application was usually limited to tracking stationary objects [53,83-86,88,89] whereas ours is on moving objects. Our future work is to develop an NN-based tracking method that is computationally light enough to enable timely tracking in the mxdR-HMDs as well as good accuracy for enabling seamless TUI integration.

    We anticipate that the newer mxdR-HMD devices with a better depth camera, such as with HoloLens 2, would allow improved stability in a depth image acquisition and thus better tracking accuracy to mxdR-HS. To our best knowledge, the quality of the depth camera of HoloLens2 has not been investigated, and as future work, we will study the potential enhancement of mxdR-HS based upon HoloLens2 by porting the current implementation. Additionally, we designed our mxdR-HS interaction schema for the user to interact with the PO and hologram using only one hand which means that only one PO needs to be tracked. Interactions from multiple POs may extend the utility of our mxdR-HS in varying applications and we also plan to test this hypothesis in future work.

## 8 Conclusion

We have presented the findings from our mxdR-HS, which is a marker-less TUI for holographic volume visualization and interaction in mxdR-HMD, using the HoloLens with its built-in depth camera. Our results show that it can function at interactive rates with accuracy that is practical for mxdR applications in the context of complex volumetric medical imaging data and with multimedia data.

## Appendix A  Results of Patch-based intensity enhancement experiment

We present the computational latency (lower is better) in Table 3 according to the 7 variations in patch sizes used for sampling the intensity values of corners in mxdR-HS. For PO#1 and PO#2, the average computational latency increased as the patch size became larger. The degree of computational latency increment was bigger on larger patch sizes as depicted in the line graph's curved slope of Fig. 9.

**Table 3** The average and standard deviation of computational latency in milliseconds for PO#1 and PO#2 with 7 patch sizes

| Patch size | PO#1 Average | PO#1 Standard deviation | PO#2 Average | PO#2 Standard deviation |
|---|---|---|---|---|
| 1×1 px | **6.8** | **2.3** | **6.6** | **3.3** |
| 3×3 px | 7.2 | **2.3** | 7.0 | **3.3** |
| 5×5 px | 8.0 | **2.3** | 7.8 | 3.4 |
| 7×7 px | 9.5 | 2.5 | 9.0 | 3.5 |
| 9×9 px | 11.2 | 2.6 | 10.7 | 3.5 |
| 11×11 px | 13.3 | 2.7 | 12.6 | 3.6 |
| 13×13 px | 16.2 | 2.9 | 15.4 | 3.9 |

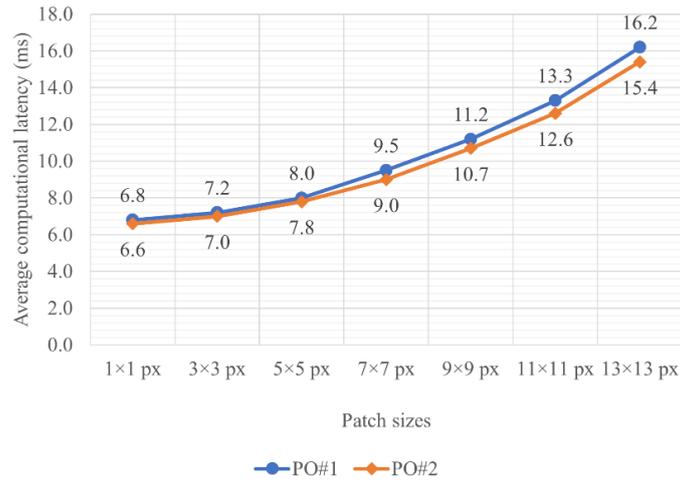

**Fig. 9** Average computational latency (y-axis) of 7 different patch sizes (x-axis)

The tracking accuracy (lower is better) of each patch size is presented in Table 4. The largest improvements of average tracking accuracy (4.5 mm on PO#1 and 4.7 mm on PO#2) were found between the single and 3×3 px patch sizes as shown in Fig. 10. The average tracking accuracy was improved when the patch sizes become larger, but the degree of improvement was smaller than what has been achieved between the single and 3×3 px patch sizes (≤ 0.5 mm on PO#1 and ≤ 0.8 mm on PO#2). The average tracking accuracy started to decrease with the patch size larger than 5×5 px on PO#1 and 11×11 px on PO#2.

**Table 4** The average and standard deviation of tracking accuracy in millimeters for PO#1 and PO#2 with 7 patch sizes

| Patch size | PO#1 | | PO#2 | |
|---|---|---|---|---|
| | Average | Standard deviation | Average | Standard deviation |
| 1×1 px | 15.9 | 6.0 | 17.2 | 6.6 |
| 3×3 px | 11.4 | 3.7 | 12.5 | 4.0 |
| 5×5 px | **10.9** | 3.2 | 11.7 | 3.7 |
| 7×7 px | **10.9** | **3.0** | 11.5 | 3.7 |
| 9×9 px | 11.0 | 2.9 | 11.4 | 3.7 |
| 11×11 px | 11.2 | 2.9 | **11.3** | **3.6** |
| 13×13 px | 11.4 | 3.0 | 11.4 | **3.6** |

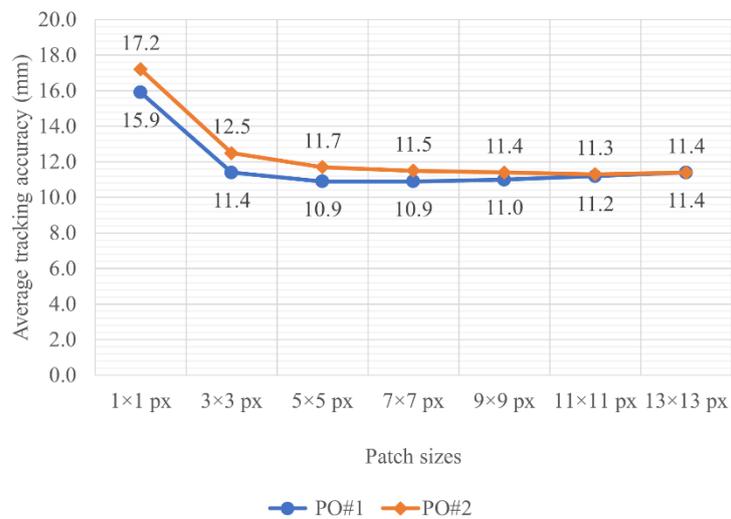

**Fig. 10** Average tracking accuracy (y-axis) of 7 different patch sizes (x-axis)